\documentclass[conference]{IEEEtran}
\IEEEoverridecommandlockouts

\usepackage{cite}
\usepackage{amsmath,amssymb,amsfonts}
\usepackage{graphicx}
\usepackage{textcomp}
\usepackage{xcolor}
\usepackage{url}

\def\BibTeX{{\rm B\kern-.05em{\sc i\kern-.025em b}\kern-.08em
    T\kern-.1667em\lower.7ex\hbox{E}\kern-.125emX}}

\makeatletter

\makeatletter
\def\ps@IEEEtitlepagestyle{%
\def\@oddhead{\parbox[t][\height][t]{\textwidth}{\small \bf
This work has been submitted to the IEEE for possible publication. Copyright may be transferred without notice, \\
after which this version may no longer be accessible.\\
\noindent
}\hfil\hbox{}}%
}

\begin{document}

\title{QUPEX: A Streamlit-Based Framework for Visualizing Quantum-Mechanical Property Spaces of Organic Molecules}

\author{
\IEEEauthorblockN{Mirtha Pillaca Quispe$^{\ast}$\textit{$^{1}$}, Alejandra Hinostroza Caldas\textit{$^{2}$}, and Leonardo Medrano Sandonas$^{\ast}$\textit{$^{3,4}$}}
\IEEEauthorblockA{$^{1}$~Independent Researcher, 01069 Dresden, Germany\\ $^{2}$~Faculty of Computer Science, TUD Dresden University of Technology, 01062 Dresden, Germany.\\ $^{3}$~Center for Advanced Systems Understanding (CASUS), Conrad-Schiedt-Straße 20, Görlitz 02826, Germany \\ $^{4}$~Helmholtz Zentrum Dresden-Rossendorf (HZDR), Bautzner Landstraße 400, Dresden 01328, Germany \\
*Emails: mirthapq@gmail.com, l.medrano-sandonas@hzdr.de}
}

\maketitle

\begin{abstract}

The growing availability of quantum-mechanical (QM) datasets offers new opportunities to investigate structure–property and property–property relationships in drug-like molecules using data-driven methods. However, interactive tools that integrate molecular visualization, statistical analysis of QM properties, clustering, and machine learning (ML) remain limited, despite their value in facilitating dataset exploration and interpretation. To address this gap, we introduce QUPEX (``QUantum Property EXplorer''), an open-source Streamlit-based framework for the interactive exploration of QM property spaces in molecular systems. 
As a demonstration, we apply QUPEX to a subset of the QM7-X dataset comprising 41,537 equilibrium molecular structures containing up to seven non-hydrogen atoms (C, N, O, S, and Cl) and their 24 associated QM properties. In addition to property-space analysis, QUPEX incorporates an XGBoost regression pipeline for predicting PBE0 dipole moments from tight-binding-derived electronic descriptors, together with model evaluation and SHAP-based interpretability functionality. By combining interactive molecular visualization, data analysis, and predictive modeling, QUPEX provides an accessible and extensible environment for investigating molecular datasets and identifying patterns in their structural and electronic properties. These capabilities can support the exploration of chemical space and inform the rational design of drug-like molecules with targeted QM properties.

\end{abstract}


\section{Introduction}

Comprehensive and systematically generated quantum-mechanical (QM) datasets of molecules and materials have become essential resources for exploring chemical space and developing machine learning (ML) methods for predicting QM properties~\cite{datarev,susml,COLEY2021133}. Prominent examples include QM7-X~\cite{hoja2021qm7x}, QM9~\cite{Ramakrishnan2014}, ANI~\cite{ani-1x}, QMugs\cite{qmugs}, SPICE~\cite{eastman2023spice}, Aquamarine\cite{medrano2024dataset}, and OMol25\cite{levine2026openmolecules2025omol25}. 
Collectively, these datasets contain extensive QM data for chemically diverse drug-like molecules, covering equilibrium geometries, non-equilibrium conformers, reaction pathways, and condensed-phase configurations. However, the scale and multidimensional nature of these data make it challenging to identify molecules, properties, and regions of chemical space relevant to a given scientific question. Making these datasets more accessible therefore requires tools that enable users to navigate molecular structures, investigate distributions and correlations across multiple properties, and map high-dimensional representations onto interpretable low-dimensional spaces\cite{medina08,VonLilienfeld2020}.

Interactive visualization provides a natural approach to addressing these challenges by enabling researchers to explore QM datasets directly and gain insights into their underlying structural and property relationships. Streamlit~\cite{streamlit} is well suited to this purpose, providing a browser-based environment for developing lightweight interactive applications and lowering the barrier to exploratory data analysis. 
It has already been adopted in various molecular tools, including Stmol~\cite{napoles2022stmol} and streamlit-molstar~\cite{molstar_streamlit2024} for three-dimensional (3D) molecular visualization, ChemCluster~\cite{chemcluster2024} and streamlit-chem~\cite{streamlit_chem2023} for chemical space exploration, and ParetoMol~\cite{yabbarov2026paretomol} and Molecular Analyzer~\cite{molecular_analyzer2024} for molecular property analysis. 
These tools demonstrate the value of graphical interfaces for exploring complex molecular data, but focus on specific tasks, such as 3D visualization, clustering, or Pareto analysis. To our knowledge, an integrated framework that combines molecular visualization, conformational and property-space analysis, and ML modeling within a single workflow remains lacking.

\begin{figure}[t]
\centering
\includegraphics[width=0.5\textwidth]{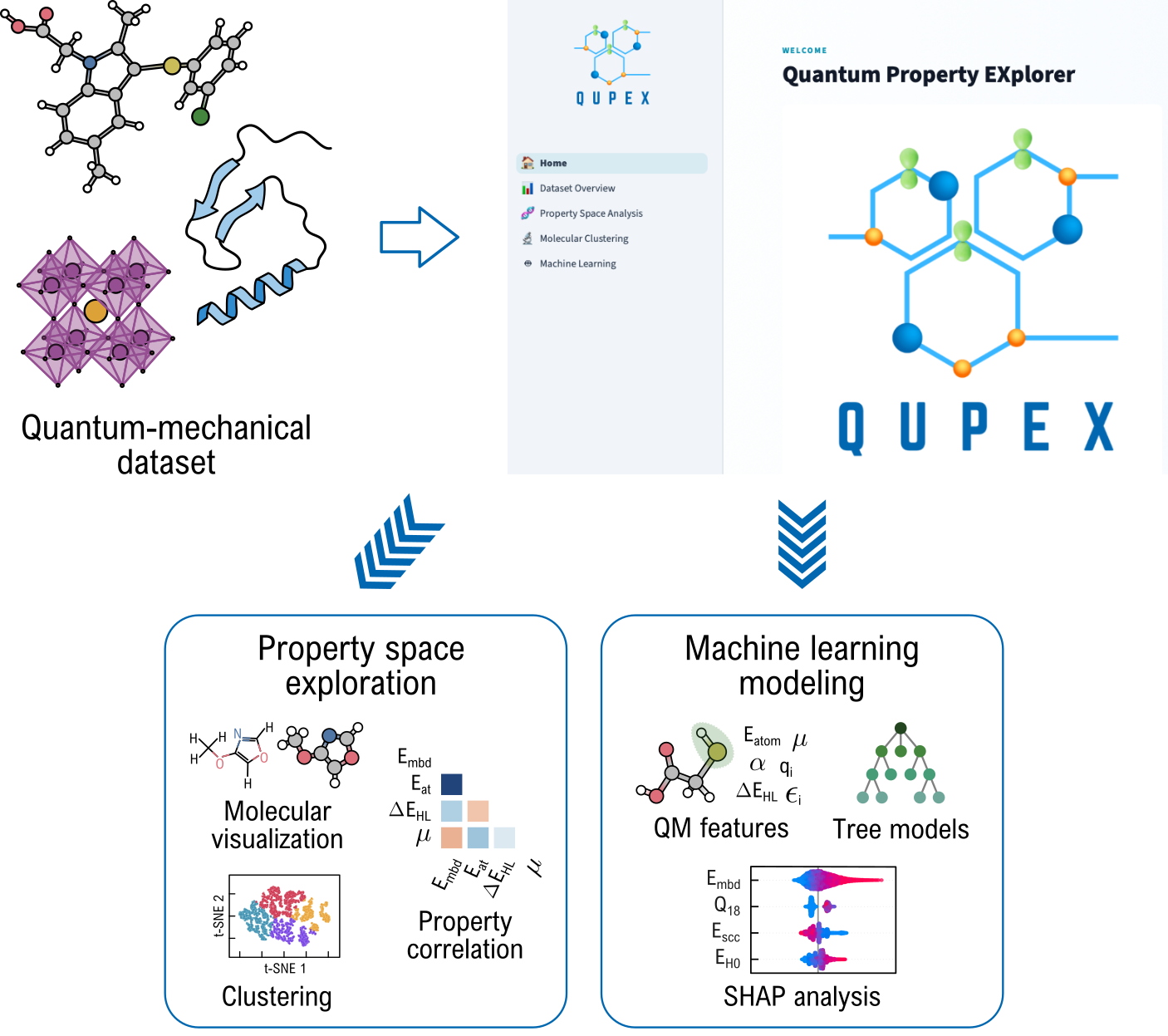}
\caption{Schematic workflow of QUPEX: a quantum-mechanical (QM) dataset is loaded and processed, enabling dataset exploration, 2D and 3D molecular visualization, Property space analysis, molecular clustering, and machine learning (ML) prediction and interpretability.}
\label{fig1}
\end{figure}

Here, we introduce QUPEX (``QUantum Property EXplorer''), a Streamlit-based framework designed to address this gap by providing an integrated environment for exploring QM datasets of drug-like molecules (Fig.~\ref{fig1}). QUPEX comprises five interconnected pages. The \texttt{Home} page allows users to load datasets from local files or Zenodo URLs. The \texttt{Dataset Overview} page provides global statistics and visualizes property distributions. The \texttt{Property Space Analysis} page enables molecule-level inspection through two-dimensional (2D) and 3D molecular visualization, alongside the exploration of QM properties. 
The \texttt{Molecular Clustering} page supports dimensionality reduction and clustering to facilitate the identification of patterns in high-dimensional property space. Finally, the \texttt{Machine Learning} page provides tools for validating predictive models and interpreting their predictions, illustrated here using XGBoost to predict the total dipole moment.

\section{Computational Methods}

\subsection{The QUPEX framework}
\label{methods}

QUPEX is a Python-based visualization framework built with Streamlit \cite{streamlit}. It consists of a landing page and four analysis pages, all of which use the shared Parquet indexes defined for a given dataset. This design avoids repeatedly scanning the source HDF5 file during user interactions. Direct access to HDF5 is reserved for retrieving detailed atomic data for a user-selected molecule or conformation. Streamlit's resource and data caches (\texttt{st.cache\_resource}, \texttt{st.cache\_data})  retain the open HDF5 handle and loaded indexes across application reruns. The five pages are described below.

\subsubsection{Home} This page initializes the workflow by allowing users to load a dataset through four sources: the benchmark QM7-X dataset hosted on Zenodo, a custom Zenodo URL, a local HDF5 file, or a file uploaded through the browser interface. For convenience, datasets downloaded from Zenodo are cached locally to enable reuse and avoid redundant downloads in subsequent runs.

\subsubsection{Dataset Overview} 
\begin{figure}[h]
\centering
\includegraphics[width=0.5\textwidth]{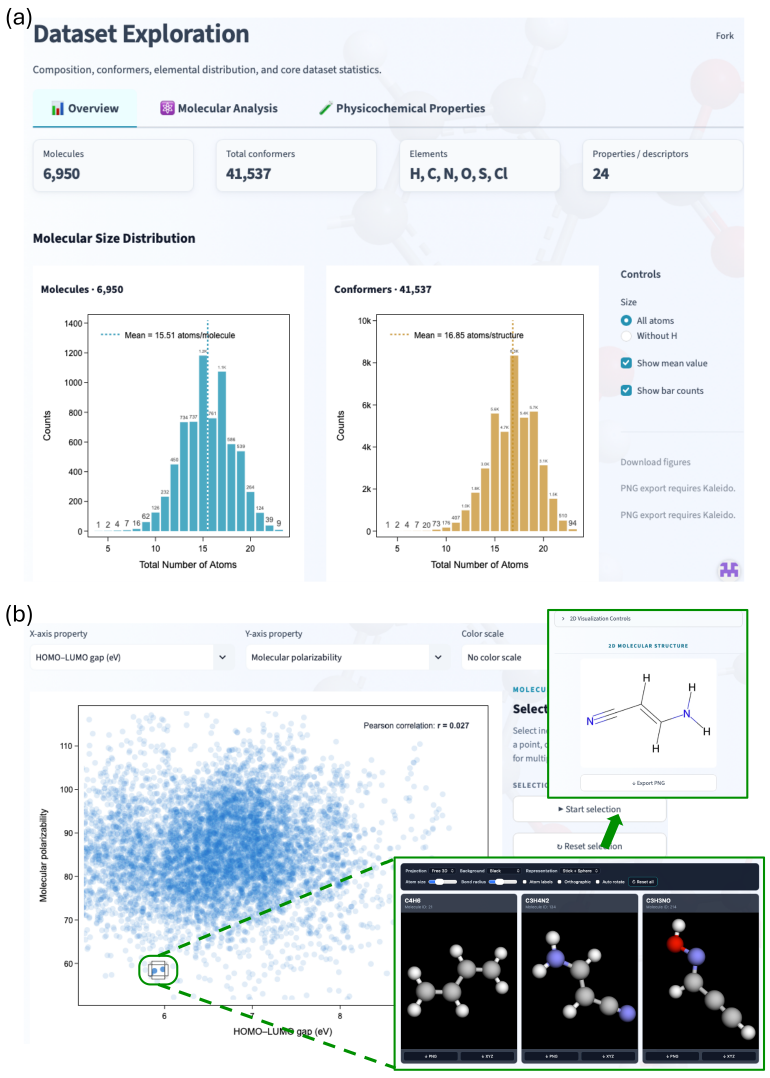}
\caption{Screenshots of the (a) \texttt{Dataset Overview} and (3) \texttt{Property Space Analysis} pages in the QUPEX Streamlit application, highlighting some of their functionalities available in each page.
}
\label{fig2}
\end{figure}

This page summarizes dataset-level statistics and property distributions through three tabs. The \emph{Overview} tab presents general dataset and conformers statistics (Fig.~\ref{fig2}(a)). The \emph{Molecular Analysis} tab describes elemental composition and molecular formulas, and provides a conformer-identifier browser that exposes the Molecule $\rightarrow$ Isomer $\rightarrow$ Conformation hierarchy. The \emph{Physicochemical Properties} tab presents a table of the available properties.

\subsubsection{Property Space Analysis} This page supports the exploration of property distributions and correlations through three tabs: \emph{Property Distributions}, \emph{Correlation Matrix} (using Pearson and Spearman methods), and \emph{Pairplots \& Molecular Structures}. The first two tabs provide statistical views of the indexed scalar properties, while the third links interactive scatter plots and pairplots to a molecular structure viewer. The viewer combines a 2D RDKit~\cite{rdkit} representation with an interactive 3D py3Dmol viewer, offering adjustable projections, molecular representations, and atom labels, together with an atom-color legend.
The page supports both single-molecule inspection and comparative analysis of multiple molecules. Users can select from the pairplots individual molecules by clicking a point, or use the box/lasso selection for multiple molecules. Selected molecules are then displayed side by side, alongside an aligned comparison of their properties (Fig.~\ref{fig2}(b)). Additionally, atom-level inspection is available by selecting an individual atom within a chosen molecule.

\begin{figure}[h]
\centering
\includegraphics[width=0.5\textwidth]{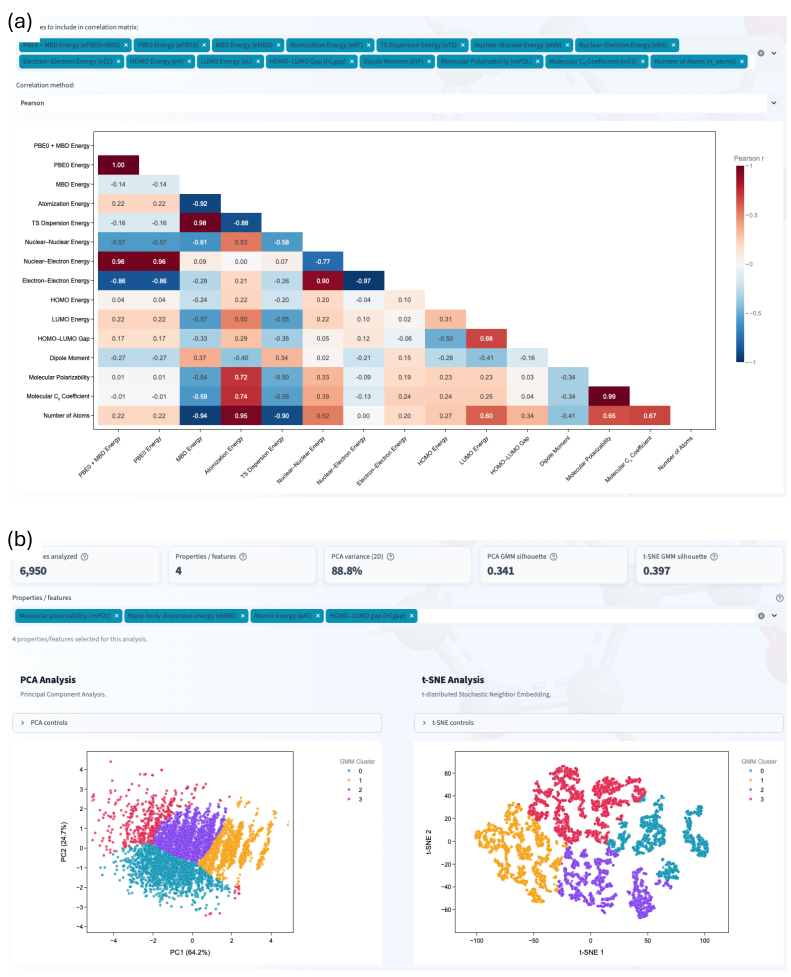}
\caption{Screenshots of the (a) \texttt{Property Space Analysis} and (b) \texttt{Molecular Clustering} pages in the QUPEX Streamlit application, showing the property correlation matrix as well as the clustering and visualization of molecules based on reduced property set, respectively.}
\label{fig3}
\end{figure}

\subsubsection{Molecular Clustering} This page enables dimensionality reduction and exploratory analysis of molecular similarity. Users select a subset of scalar properties and apply Principal Component Analysis (PCA) or t-distributed Stochastic Neighbor Embedding (t-SNE), as implemented in \texttt{modules/chemical\_space.py}. The resulting representations are used for Gaussian Mixture Model (GMM) clustering with 2--10 components (Fig. \ref{fig3}(b)). The PCA and t-SNE GMMs are fitted independently, each directly on its own 2D projection. Cluster quality is evaluated using the silhouette score, while GMM confidence ellipses are projected into PCA space. The page also compares PCA- and t-SNE-based cluster assignments using the adjusted Rand index. 
The underlying module additionally implements K-means clustering through \texttt{run\_kmeans}; however, this functionality is not currently exposed in the user interface, which provides only the GMM-based workflow. Because the results depend on the selected properties, dimensionality-reduction method, and clustering parameters, this page is intended for exploratory analysis rather than for establishing a single, definitive partitioning of chemical space.

\subsubsection{Machine Learning} 
\begin{figure*}[h]
\centering
\includegraphics[width=\textwidth]{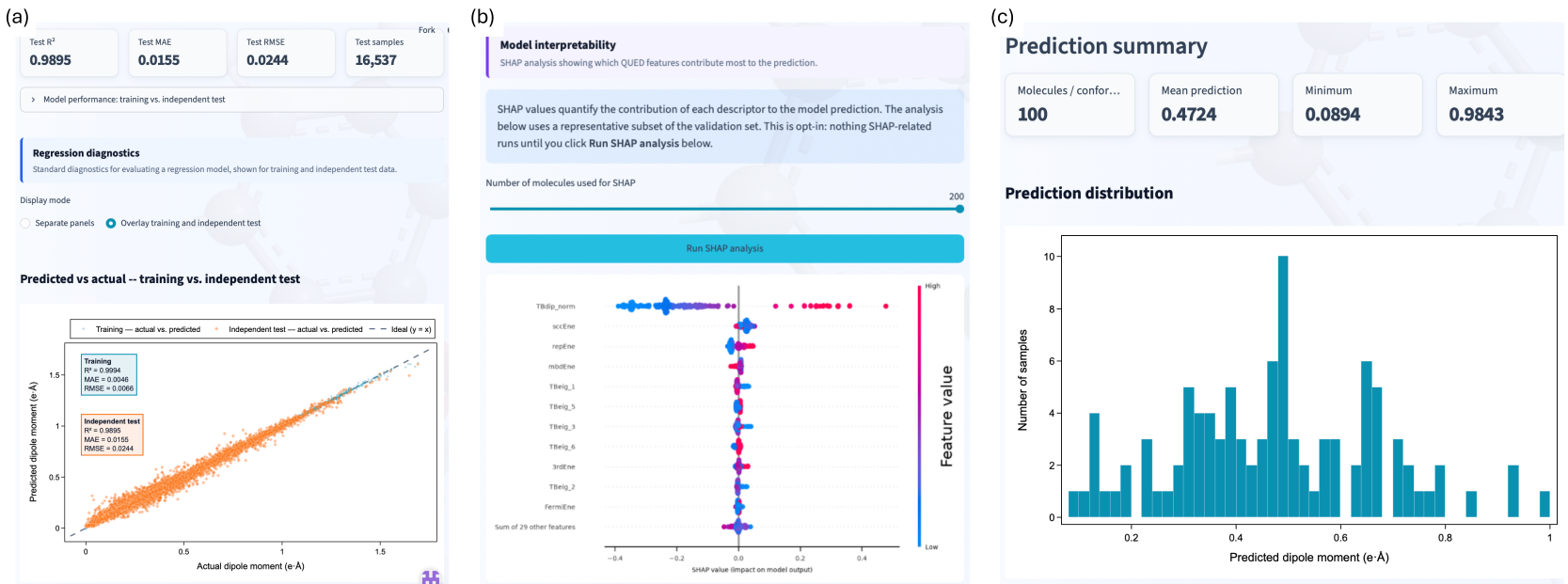}
\caption{Screenshots showing the main features of the \texttt{Machine Learning} page in the QUPEX Streamlit application: (a) statistical metrics and evaluation of the trained XGBoost model on the training and test sets using standard regression diagnostics, (b) SHAP-based model interpretability, and (c) statistical metrics and the distribution of predicted dipole moments.}
\label{fig4}
\end{figure*}

This page demonstrates property prediction using a pretrained ML model and supports inference on new datasets (Fig. \ref{fig4}). The model is a scikit-learn~\cite{pedregosa2011scikit} pipeline comprising a \texttt{StandardScaler} and an \texttt{XGBRegressor}~\cite{chen2016xgboost}. The page provides two tabs. The \emph{Model validation} tab evaluates the pretrained model on a bundled test set and provides model interpretability analysis using SHapley Additive exPlanations (SHAP)~\cite{lundberg2017unified}. The \emph{New-data prediction} tab accepts HDF5 or CSV datasets containing compatible electronic descriptors and applies the pretrained model to generate predictions. Further details of the model and its performance are provided in Sec.~\ref{sec:ml}.

\subsection{QM7-X property space}

To demonstrate the capabilities of the QUPEX visualization framework, we consider the widely used QM7-X dataset~\cite{hoja2021qm7x}. QM7-X was constructed by extensively sampling an initial set of 6,950 small organic molecules, including their isomers and conformers, and generating 100 non-equilibrium structural variations for each obtained equilibrium structure. The resulting dataset comprises approximately 4.2 million structures containing up to seven heavy atoms (C, N, O, S, and Cl). For each structure, 42 molecular and atomic QM properties were computed. Most of them were calculated at the PBE0+MBD level of theory~\cite{pbe0a,adamo1999,tkatchenko12}, using tightly converged numeric atom-centered basis sets~\cite{havu2009efficient}, as implemented in the \texttt{FHI-aims} package~\cite{fhiaims2026}.
In this work, we focus on the property space spanned by a subset of 41,537 equilibrium structures and a selected set of 24 QM properties, comprising 19 molecular and five atomic properties. Examples of the investigated properties are listed in Table~\ref{table1}.

\begin{table}[t]
\caption{Excerpt of the reference table of properties associated to QM7-X equilibrium molecules.}
\label{table1}
\centering
\begin{tabular}{lll}
\hline
\textbf{Code} & \textbf{Meaning} & \textbf{Units} \\
\hline
atXYZ      & Cartesian coordinates        & \AA \\
atNUM      & Atomic numbers        & -- \\
eAT        & Atomization energy                & eV \\
eMBD       & Many-body dispersion energy  & eV \\
eXC       & Exchange–correlation energy  & eV \\
DIP        & Dipole moment                & e$\cdot$\r{A} \\
HLgap      & HOMO--LUMO gap               & eV \\
mC6        & Molecular C6 coefficient     & a.u. \\
mPOL       & Molecular polarizability     & a$_0^3$ \\
hCHG       & Hirshfeld atomic charges    & e \\
\hline
\end{tabular}
\end{table}

\subsection{Machine learning modeling}
\label{sec:ml}

Building on our recently developed QUED framework (``QUantum Electronic Descriptors'')\cite{qued}, we trained an ML model to predict the PBE0 reference dipole moment (DIP) of QM7-X molecules from electronic descriptors derived using the semi-empirical third-order density-functional tight-binding method (DFTB3) with many-body dispersion (MBD) corrections\cite{hourahine2020}. The 40-dimensional descriptor comprises nine global and electronic quantities (FermiEne, BandEne, NumElec, h0Ene, sccEne, 3rdEne, repEne, mbdEne, and TBdip), eight tight-binding orbital eigenvalues (TBeig), and 23 Mulliken atomic charges (TBchg). The atomic charges are zero-padded to a fixed length of 23, corresponding to the maximum number of atoms in the QM7-X molecules considered here. This representation enables the use of a fixed-dimensional input vector across molecules of different sizes.

The bundled model (\texttt{models/model-qued-dip.pkl}) is a scikit-learn pipeline comprising a \texttt{StandardScaler} and an \texttt{XGBRegressor}~\cite{chen2016xgboost}. It was developed using the selected subset of 41,537 equilibrium structures, divided into a training set of 25,000 molecules and a test set of 16,537 molecules. Training samples were selected using farthest point sampling (FPS)~\cite{yuze2025farthest}, while the remaining samples were assigned to the test set. Hyperparameter optimization was performed using Optuna~\cite{optuna_2019}, a Bayesian optimization framework. Each optimization trial employed five-fold cross-validation, with the objective of minimizing the root mean square error (RMSE). The optimization procedure comprised 100 trials.
QUPEX also implements an interpretability module based on SHAP \texttt{TreeExplainer}, which generates beeswarm summary plots for user-selected subsets of 25--200 test molecules. In addition, an inference interface accepts new HDF5 or CSV datasets containing compatible electronic descriptors and applies the pretrained model to generate predictions. These capabilities are documented as implemented features of QUPEX

\section{Results \& Discussion}
\label{sec:discussion}

\subsection{Framework initialization}

The \texttt{Home} page of the QUPEX visualization framework provides three options for loading a dataset: selecting a local file, uploading a file through the browser, or specifying a Zenodo URL. The navigation sidebar provides access to four analysis pages: \texttt{Dataset Overview}, \texttt{Property Space Analysis}, \texttt{Molecular Clustering}, and \texttt{Machine Learning}. Together, these pages establish an integrated workflow that progresses from dataset-level structural and compositional statistics to interactive molecular visualization and property correlation, followed by exploratory clustering in property space and supervised property prediction with model interpretability.

Throughout this work, we distinguish three levels of evidence: (i) \emph{implemented functionality}, which can be verified directly from the application's source code; (ii) \emph{exploratory analyses}, which the application can perform for a specified user-selected configuration but which are not presented as fixed scientific findings; and (iii) \emph{validated, reproducible results}, for which the reported values were independently regenerated using the project's code and data during preparation of this work. The application and its source code are publicly available on GitHub (https://github.com/Mirtha-Pillaca/QUPEX), enabling independent inspection of the implementation and verification of the reported results.

As a proof-of-concept demonstration, we applied QUPEX to the equilibrium structures in the QM7-X dataset~\cite{hoja2021qm7x}. The QM data for the 41,537 equilibrium structures considered here are stored in the HDF5 file \texttt{eq-QM7X.hdf5}. Rather than loading the entire HDF5 file into memory, QUPEX uses an indexing pipeline implemented in \texttt{modules/prepare\_indexes.py} and \texttt{modules/loader.py} to generate two cached Parquet tables: a molecule-level index with one row per optimized (unique) molecule and a conformer-level index with one row per geometry. These indexes contain scalar QM7-X properties, elemental atom counts, and parsed geometry identifiers that encode the Molecule $\rightarrow$ Isomer $\rightarrow$ Conformation hierarchy (e.g., \texttt{Geom-m2682-i4-c3-opt}).
For the QM7-X dataset, the resulting indexes cover 6,950 unique molecules and 41,537 geometry records. Across the 6,950 molecules, QUPEX identifies 107,764 atoms, comprising 60,495 H, 34,686 C, 6,438 N, 5,830 O, 292 S, and 23 Cl atoms. These molecules represent 182 unique molecular formulas. QUPEX also provides a curated reference table of 24 of the 42 QM properties available in QM7-X, with human-readable names and units. The complete set of properties is described in Ref.~\cite{hoja2021qm7x}, while a representative subset of the QUPEX reference table is presented in Table~\ref{table1}.

Beyond the specific QM7-X file used in this demonstration, QUPEX implements a dataset-loading and indexing pipeline designed to accommodate other compatible datasets. The pipeline inspects the input file's structure, including conformer-identifier conventions and property-name signatures, to determine whether it matches the QM7-X format or should be handled as a generic custom HDF5 dataset. It also stores the cached Parquet indexes under a stable, dataset-specific identifier, preventing index collisions between different source files. Consequently, QUPEX can, in principle, index and analyze other datasets that follow a compatible, QM7-X-like structure, rather than being restricted to the specific file used during development.

\subsection{Assessing QUPEX capabilities}

The exploration of high-dimensional property spaces of drug-like molecules is challenging but offers considerable potential for advancing multi-property optimization and molecular design~\cite{medrano23fod,qalpa}. Understanding correlations between molecular properties can help identify distinct molecules that share desirable combinations of electronic features, thereby narrowing the pool of candidates for a given technological application. QUPEX facilitates this exploration by providing correlation analyses across multiple molecular and atomic properties, together with interactive molecular visualization. As illustrated in Fig.~\ref{fig2}(b), users can select molecules directly from correlation plots, inspect their structures, and compare their atomic features side by side. This integrated analysis enables users to investigate how differences in molecular structure and chemical configuration are associated with variations in QM properties, providing a basis for identifying potential chemical design principles.
QUPEX also provides correlation matrices for user-selected molecular properties, calculated using either the Pearson or Spearman correlation coefficient (Fig.~\ref{fig3}(a)). These matrices offer a global overview of the relationships between properties across the dataset and help users identify potentially relevant property combinations for further investigation. Whereas Pearson correlation captures linear relationships, Spearman correlation assesses monotonic relationships based on ranked values, providing complementary perspectives on the property space.

QUPEX further supports the identification of molecular clusters within a user-defined property space. To facilitate the exploration of the high-dimensional QM7-X dataset, the implemented workflow first standardizes the selected scalar properties and applies Principal Component Analysis (PCA, \texttt{modules.chemical\_space.analyze\_pca}) to obtain a two-dimensional representation using the first two principal components (Fig.~\ref{fig3}(b)). Alternatively, users can apply t-distributed Stochastic Neighbor Embedding (t-SNE) to generate a low-dimensional representation. The resulting embeddings can then be analyzed using Gaussian Mixture Models (GMMs) or K-means clustering to identify groups of molecules with similar property profiles. The interface allows users to adjust the dimensionality-reduction and clustering parameters interactively, making this functionality an exploratory tool for investigating patterns in property space rather than imposing a single, fixed partitioning of the dataset.

The \texttt{Machine Learning} page integrates predictive modeling into the QUPEX workflow. It provides an interface for evaluating a pretrained XGBoost regression model that predicts the PBE0 reference dipole moment of small molecules from 40-dimensional electronic descriptors derived using DFTB3 with MBD correction. The model can be evaluated on the training and test sets defined in this work, containing 25,000 and 16,537 structures, respectively, using the validation functionality implemented in \texttt{pages/4\_Machine\_Learning.py}. The corresponding test-set data are stored in \texttt{models/qm7x-eq-test.h5}. As shown in Fig.~\ref{fig4}(a), the model achieves (R$^2$=0.999), a mean absolute error (MAE) of 0.005 e$\cdot$\r{A}, and a root mean square error (RMSE) of 0.007 e$\cdot$\r{A} on the training set. On the test set, it achieves R$^2$=0.989, an MAE of 0.015 e$\cdot$\r{A}, and an RMSE of 0.024 e$\cdot$\r{A}. The optimized XGBoost hyperparameters are stored in the serialized model and comprise \texttt{reg\_lambda} = 8.0, \texttt{reg\_alpha} = 0.01, \texttt{learning\_rate} = 0.012, \texttt{n\_estimators} = 3000, and \texttt{max\_depth} = 15.

To support model interpretation, QUPEX also incorporates a module based on SHAP that allows users to examine the contributions of individual DFTB-derived features to the predicted dipole moments (Fig.~\ref{fig4}(b)). Users can select a subset of test samples and generate SHAP summary plots to investigate the relative importance and distribution of feature contributions. The interactive interface focuses on test-set samples to facilitate the interpretation of model behavior on previously unseen data. Finally, QUPEX provides an inference interface through which users can supply HDF5 or CSV files containing compatible electronic descriptors for molecules outside the QM7-X dataset and obtain dipole-moment predictions (Fig.~\ref{fig4}(c)). This functionality offers a potential route to screening molecular candidates when dipole moment is a relevant selection criterion, provided that the input descriptors are compatible with the model and the predictions fall within its domain of applicability.

\section{Conclusions}

In this work, we presented QUPEX (``QUantum Property EXplorer''), an open-source Streamlit-based framework designed to facilitate the interactive exploration and analysis of molecular QM datasets. By integrating 2D/3D molecular visualization, property distributions and correlations, dimensionality reduction, clustering, and ML functionalities, QUPEX provides a unified environment for examining molecular structures and their associated properties. 
Its application to the QM7-X dataset illustrates how these capabilities can support the exploration of chemical space and the analysis of structure–property and property–property relationships. Such analyses can help assess established chemical and physical principles and identify patterns that may inform the rational design of molecules with targeted properties.
The integrated XGBoost pipeline further illustrates the potential of combining semi-empirical electronic descriptors with predictive modeling and interpretability tools such as SHAP analysis. The use of indexed data and caching mechanisms also supports efficient interaction with datasets containing thousands of molecular configurations. 
QUPEX is publicly available, together with its source code and associated data, enabling reproducibility and community-driven extensions. Future developments could expand its compatibility with additional QM datasets, descriptors, and ML models, further strengthening its role as a flexible tool for molecular data exploration and analysis.

\section{Data Availability}

The source code of the Streamlit-based QUPEX visualization framework is publicly available on GitHub at https://github.com/Mirtha-Pillaca/QUPEX, together with instructions for installation and use. The subset of the QM7-X dataset containing the structural and QM property data for equilibrium structures is available in HDF5 format from Zenodo at https://zenodo.org/records/22893015. Data files defining the training and test sets used to develop the XGBoost regression model for dipole-moment prediction, together with the trained model (\texttt{qued-model\_dip.pkl}), are also provided. QUPEX application can be accessed online at https://qupex-app.streamlit.app. 

\bibliographystyle{IEEEtran}
\bibliography{references}

\end{document}